\documentclass{article}
\usepackage[utf8]{inputenc}
\usepackage{amsmath}
\usepackage{amssymb}
\usepackage{authblk}
\usepackage{bigfoot}
\usepackage{tikz}
\usetikzlibrary{automata}
\usetikzlibrary{arrows}
\usepackage{nicematrix}
\usepackage{float}
\usepackage[backend=biber,style=alphabetic]{biblatex}

\usepackage{listings}
\usepackage{xcolor}
\usepackage{comment}

\lstdefinestyle{mystyle}{
    backgroundcolor=\color{backcolour},   
    commentstyle=\color{codegreen},
    keywordstyle=\color{magenta},
    numberstyle=\tiny\color{codegray},
    stringstyle=\color{codepurple},
    basicstyle=\ttfamily\footnotesize,
    breakatwhitespace=false,         
    breaklines=true,                 
    captionpos=b,                    
    keepspaces=true,                 
    numbers=left,                    
    numbersep=5pt,                  
    showspaces=false,                
    showstringspaces=false,
    showtabs=false,                  
    tabsize=2
}

\definecolor{codegreen}{rgb}{0,0.6,0}
\definecolor{codegray}{rgb}{0.5,0.5,0.5}
\definecolor{codepurple}{rgb}{0.58,0,0.82}
\definecolor{backcolour}{rgb}{0.95,0.95,0.92}

\title{A New Outlook on the Profitability of Rogue Mining Strategies in the Bitcoin Network}
\author[1]{Pantelis Tassopoulos}
\author[2]{Yorgos Protonotarios}
\affil[1]{Department of Mathematics, Imperial College London}
\affil[2]{Department of Mathematics, University College London}

\date{July 2022}

\begin{document}

\maketitle
\section{Abstract}
Many of the recent works on the profitability of  rogue mining strategies hinge on a parameter called gamma ($\gamma$) that measures the proportion of the honest network attracted by the attacker to mine on top of his fork. These works, see \cite{https://doi.org/10.48550/arxiv.1808.01041} and \cite{https://doi.org/10.48550/arxiv.1805.08281}, have surmised conclusions based on premises that erroneously treat $\gamma$ to be constant. In this paper, we treat $\gamma$ as a stochastic process and attempt to find its distribution through a Markov analysis. We begin by making strong assumptions on gamma's behaviour and proceed to translate them mathematically in order to apply them in a Markov setting. The aforementioned is executed in two separate occasions for two different models. Furthermore, we model the Bitcoin network and numerically derive a limiting distribution whereby the relative accuracy of our models is tested through a likelihood analysis. Finally, we conclude that even with control of 20\% of the total hashrate, honest mining is the strongly dominant strategy.

\section{Introduction}
In this section we aim to explain various fundamental concepts used in our research below and how they are interrelated. To begin with, a description of the three rogue mining strategies investigated in this paper is warranted. Furthermore, the concept of a difficulty adjustment which is exploited in these strategies is of paramount importance. To accompany the aforementioned, it is also essential to explain the importance of \(gamma\) in these strategies.\\
\\Mining a block entails "finding" the target cryptographic hash of the block. The target hash is a hash that begins with a predetermined number of zeros. A miner concatenates the version of the current Bitcoin software, the timestamp of the block, the root of its' transaction's merkle tree, the difficulty target and the nonce and inputs them in the SHA-256 hashing function to obtain an output. The nonce is the only variable quantity out of these six elements. Hence, the miner only varies the nonce and inputs it in the SHA-256 hashing function in the hopes of obtaining the target hash. "Obtaining the target hash" does not mean having the identical hash being output from the SHA-256 algorithm; it means obtaining a hash that has the same or more leading zeros. The difficulty is defined as the number of leading zeros contained in the target hash. The Bitcoin network demands a block be mined in 10 minutes and after 2016 blocks the network evaluates whether these blocks have been approximately mined in 20,160 minutes. The difficulty adjustment primarily depends on the number of miners or more precisely the hashing power of the sum of all miners. If the totality of the miners have taken more time to do so than the network adjusts the difficulty by reducing the number of leading zeros and if not the analogous occurs.\\
\\In this paper we make use of three rogue strategies. These are: Selfish Mining (SM), Least-Stubborn Mining (LSM) and Equal Fork Stubborn Mining (EFSM). The two latter ones are slight modifications of the popular Selfish Mining attack. SM is a strategy that targets the difficulty adjustment of the protocol by invalidating "honest" blocks through broadcasting a chain of secretly mined blocks which results in slowing down the network and hence the difficulty becomes easier even though the hashing power has not changed. Henceforth, the revenue of a miner per unit time increases. EFSM and LSM differentiate from SM only in terms of the timing on when the secret chain is revealed as well as the fact that the miner also has the choice to strategically reveal blocks instead of the entire chain when it comes to EFSM and LSM. For a complete a description of the strategies we refer the reader to \cite{https://doi.org/10.48550/arxiv.1808.01041} and \cite{https://doi.org/10.48550/arxiv.1805.08281}.\\
\\The parameter $\gamma$ appears when a fork between a rogue chain and an honest chain occurs (see \cite{nayak2016stubborn}). In such scenarios, there exists a fraction of honest miners (in other words $\gamma$) that mine on top of the rogue chain. This parameter is instrumental in the investigation of the optimality of rogue mining strategies; thence, we investigate its behaviour in this paper.

\section{Bitcoin Network}
\noindent We are going to outline and motivate the construction of a Bitcoin network, mirroring many aspects of the existing network. This will be used as a proxy to test the relative fitness of our analytical Markov models for the distribution of \(\gamma\).\\

\noindent Tools from graph theory will be used to construct a numerical model that will be used to stochastically simulate the Bitcoin network using a series of increasing times \(\tau\) (see code excerpt 2), that represent the real times since the first instance where two nodes ping the network and the response in terms of \(\gamma\) is recorded and stored in an array. By sampling from such a sequence, of times, we obtain a time series of values of \(\gamma\), whence we compute the transition probabilities between optimal mining strategies in the Markov chain model for gamma, and as a by product, we get its limiting distribution (see code excerpt 1).\\

\noindent The nodes in the network are meant to correspond to mining pools across the world, each in a specific continent. The amount of nodes in each continent is determined by the fraction of the hash rate contributed to the Bitcoin network by each continent respectfully \cite{https://doi.org/10.48550/arxiv.1901.09777} (see code excerpt 7).

\subsection{Construction}

\noindent The Bitcoin network at any given time \(t \geq 0 \) will be modelled by a weighted graph \[\mathcal{G}_{t} = (V_{t}, E_{t})\] with \(V_{t} = \{1,2,3\dots,100\}\) vertices corresponding to nodes on the network, and \(E_{t} = \left\{\{i,j\}| \forall 1 \leq i < j \leq 100\right\} \) edges, with a (stochastic - its precise nature will be explained later) weight function \[\mathcal{W}_{t}:E_{t} \rightarrow \mathbb{R} \] that measures the latency of nodes between themselves in microseconds.\\

\noindent Key assumptions on the latencies between the nodes that will be explored further below are:

\begin{itemize}
    \item network topology
    \item historical latency
    \item skew normality of latency distribution
    \item time separation between the measurements
\end{itemize}

\noindent To account for geographical separation between the nodes, values for the mean latencies between continents in the Bitcoin network in 2019 (see \cite{https://doi.org/10.48550/arxiv.1901.09777}) were used in the weights of the graph \(\mathcal{G}_{t}\).\\

\noindent Additionally, the topology of the network, that is the combinatorial properties of the underlying graph used to model the network itself (see \cite{trudeau2013introduction}, p.76), will have an impact on the connectivity of the nodes therein. More specifically, the notion of eigenvalue centrality plays a crucial role in determining the weights of the network. The utility of this metric lies in that heuristically, nodes with high centrality are connected to proportionately more nodes with high scores \cite{newman2008mathematics}. To make this mathematically precise, one takes the adjacency matrix of the graph upon initialisation of the graph's weights in the simulation at a given time; some of the weighs may take the value \(1E7\), which is to be interpreted that the connection between the nodes is non existent at that moment. Then, one computes the adjacency matrix of the graph, defined by: 

\[
\bf{A}_{ij}= \left\{
\begin{array}{ll}
      1 \text{ if  } \mathcal{W}({i,j)}< 1E7\\
      0 \text{ otherwise}\\
\end{array} 
\right. 
\]
for \(\{i,j\}\in V\). This is then used to compute the centrality score vector \(\bf{\Omega}\) which satisfies: 
\[
\lambda\bf{\Omega} = A.\bf{\Omega}
\]

\noindent which satisfies \(\bf{\Omega}(i)\geq 0\) for all \(i \) vertices in the graph and \[\displaystyle \sum_{i\in V}\bf{\Omega}(i) = 1\]

\noindent We remark that its existence is guaranteed by the Perron - Frobenius Theorem\cite{newman2008mathematics}. With regards to modelling latencies on the network, we observe that on a mining network following the Bitcoin protocol, the latencies follow a multimodal distribution (see \cite{gencer2018decentralization}, figure 3). For this reason, it will be assumed that the weights of the network will follow a skew-normal distribution with shape parameter \(\alpha\) depending on the combined eigenvector centrality of the nodes comprising an edge.\\

\subsection{Modelling Assumptions}
Before diving into the modeling assumptions, it is important to state that mining is a Markov process, see \cite{https://doi.org/10.48550/arxiv.2003.00001}. Let $\gamma_n$ for \(n \in \mathbb{N}\) represent the process modelling \(\gamma\) in discrete time, and consider the modified stochastic process indexed by \(\mathbb{N}\): 

\[\displaystyle X_n =  \bigcup_{k \in T} \mathbf{1}_{A_{k}}(\gamma_n) : \mathbb{N} \rightarrow \Xi\]

\noindent where \[\mathbf{1}_{A_{k}}(x) = 
\begin{cases}
  A_k & \text{if $x \in A_k$,} \\
  \emptyset & \text{otherwise}
\end{cases}
\]

\noindent and $T \subset \mathbb{N},$ $|T|<\infty$, and $\Xi = \{A_{k} : k \in T\}$ such that \[A_{i} \bigcap A_{j} = \emptyset \text{\space \space} \forall i \neq j \in T\] and $\bigcup _{k \in T} A_{k} = [0, 1]$ For our following model to predict transition probabilities between strategies we require to satisfy the following ideas:\\

1. The probability that $\gamma$ jumps to an interval that is further away to be smaller than the probability of it jumping to interval nearby\\

2. The probability that $\gamma$ jumps to an interval with greater length to be greater than the probability that it jumps to an interval of smaller length.\\
\\The intuitive idea behind the above assumptions is the following. As mentioned previously $\gamma$ represents the proportion of people that follow our chain. We want the process of say a change of $\gamma = 0.2$ to $\gamma = 0.21$ to be more probable than a change from $\gamma = 0.2$ to $\gamma = 0.6$. In deed, it seems quite improbable that 20\% of people following our chain turn to 60\% compare to 21\%. Furthermore, since we give $\gamma$ a range rather than a fixed value in the Markov models, it also makes sense that if we jump to greater range of $\gamma$ we are giving ourselves more leeway than if we confined ourselves to a very small one. The above assumptions can be mathematically stated as:
\[
\mathbb{P}(X_{n_1} = [x_{1},x_{2}] | X_{n-1} = [y_{1},y_{2}]) \leq \mathbb{P}(X_{n_2} = [x_{3},x_{4}] | X_{n-1} = [y_{1},y_{2}]),\]
\noindent if 

\[ d_{p}([x_{3},x_{4}],[y_{1}, y_{2}]) \leq d_{p}([x_{1}, x_{2}],[y_{1}, y_{2}]) \text{, \space and \space} d(x_{1}, x_{2}) = d(x_{3}, x_{4})
\]
\\
Where $d_{p}(X,Y)$ is a metric defined in the following way:
\[    d_{p}: \mathcal{P}([0,1]) \times \mathcal{P}([0,1]) \longrightarrow \mathbb{N}\]
\[
d_{p}([x_{1}, x_{2}],[y_{1}, y_{2}]) = d\Big(x_{1}+\frac{x_{2}-x_{1}}{2}, y_{1}+\frac{y_{2}-y_{1}}{2}\Big)
\]Where  $d(.,.)$ is the standard metric. Moreover, we also require
\[
\mathbb{P}(X_{n_1} = [x_{1},x_{2}] | X_{n-1} = [y_{1},y_{2}]) \leq \mathbb{P}(X_{n_2} = [x_{3},x_{4}] | X_{n-1} = [y_{1},y_{2}]),\] 
\noindent if 
\[
d_{p}([x_{3},x_{4}],[y_{1}, y_{2}])  = d_{p}([x_{1}, x_{2}],[y_{1}, y_{2}]) \text{, \space and \space} d(x_{1}, x_{2}) \leq d(x_{3}, x_{4})
\] 

\section{Exploration of Models}
\subsection{1st Markov Model}
Based on the above, the following probability model will be used to compute transition probabilities:
\begin{equation}
\mathbb{P}(X_{n} =[x_{1},x_{2}] | X_{n-1} = [y_{1},y_{2}]) = \frac{\displaystyle \int_{x_{1}}^{x_{2}} 1 - d_p([x_{1}, x_{2}],[y_{1},y_{2}]) \;\mathrm{d}x}{\displaystyle\sum_{\xi  \in \Xi}\displaystyle \int_{x_{1}}^{x_{2}} 1 - d_p([x_{1}, x_{2}],\xi) \;\mathrm{d}x}
\end{equation}
The interval $[0,1]$ is partitioned into disjoint intervals as will be explained below which is represented by the set $\Xi$; the denominator is a sum over all such disjoint intervals.\\
Fixing the collection of nodes applying this strategy at an hashrate of 20\%, according to \cite{https://doi.org/10.48550/arxiv.1808.01041}, $\gamma$ is partitioned in the following way:
\[
\Xi = \{[0, 0.675], [0.675, 0.76], [0.76, 0.761], [0.761, 1]\}
\]
The set $\Xi$ as mentioned previously is the set encapsulating the way $\gamma$ is partitioned in correspondence with the mining strategies. From left to right we have: honest mining (HM), selfish mining (SM), Lead-Stubborn mining (LSM), Equal Fork Stubborn mining (EFSM). We will compute the transition probabilities. The same process is analogously applied for all other states.\\
\\Representing the results in a transition matrix, we obtain
\begin{center}
$\mathbf{P_{1}}$ = 
\[
\begin{bmatrix}
& 0.81 & 0.06 & 0 & 0.13\\
& 0.59 & 0.12 & 0.01 & 0.28\\
& 0.57 & 0.12 & 0 & 0.31\\
& 0.5 & 0.11 & 0 & 0.39 \\
\end{bmatrix}
\]
\end{center}
The chain is irreducible therefore the limiting distribution $\mathbf{\pi}$ can be obtained by solving the following equation\\
\[
\mathbf{\pi} = \bf{\pi}\mathbf{P_{1}}
\]
The matrix $\mathbf{P_{1}}$ has rank equal to $3$ which signifies that the solution has a dependency on one variable. This variable can be chosen to be unique since we require $\sum_{i = 1}^{4} \mathbf{\pi}_{i} = 1$. Solving the above equation and taking into consideration the aforementioned we obtain the limiting distribution where each term is rounded to significant figures:
\[
\mathbf{\pi_1} = \Big(0.73 \text{ \space  \space} 0.08 \text{ \space  \space} 0\text{ \space  \space} 0.19 \Big)\\
\]

\begin{center}
\begin{tikzpicture}[->, >=stealth', auto, semithick, node distance=4cm]
\tikzstyle{every state}=[fill=white,draw=black,thick,text=black,scale=1.2]
\node[state]    (HM)                     {$HM$};
\node[state]    (SM)[above right of=HM]   {$SM$};
\node[state]    (LSM)[below right of=HM]   {$LSM$};
\node[state]    (EFSM)[below right of=SM]   {$EFSM$};
\path
(HM) edge[loop left]     node{$0.81$}         (HM)
    edge[bend left]     node{$0.06$}     (SM)
    edge[bend left,below]      node{$0$}      (EFSM)
    edge[bend left, below]    node{$0.13$}      (LSM)
(SM) edge[bend right]    node{$0.28$}           (EFSM)
    edge[bend left]     node{$0.59$}              (HM)
    edge[bend left]     node{$0.01$}              (LSM)
    edge[loop above]    node{$0.12$}              (SM)
(LSM) edge[bend right]    node{$0.31$}           (EFSM)
    edge[bend left]     node{$0.57$}              (HM)
    edge[loop below]    node{$0$}              (LSM)
    edge[bend left]     node{$0.12$}              (SM)
(EFSM) edge[loop right]    node{$0.39$}     (EFSM)
    edge[bend right,right]    node{$0.11$}(SM)
    edge[bend right]     node{$0$}      (LSM)
    edge[bend left,above]     node{$0.5$}   (HM);
\end{tikzpicture}
\end{center}

\subsection{2nd Markov model}

\noindent Upon exploring the first Markov model, we proceed with another candidate for the distribution of gamma. This will be motivated by choice of 'Gaussian', or squared exponential kernel \[
\kappa(x,y) = \exp\left(-\frac{1}{2}\left(\frac{x-y}{l}\right)^2
\right)
\]

\noindent where \(l = \frac{1}{4}\) is chosen to be the characteristic length scale of the process \(\gamma_{t}\).

\begin{figure}
\label{fig: exp}
\includegraphics[width = \textwidth]{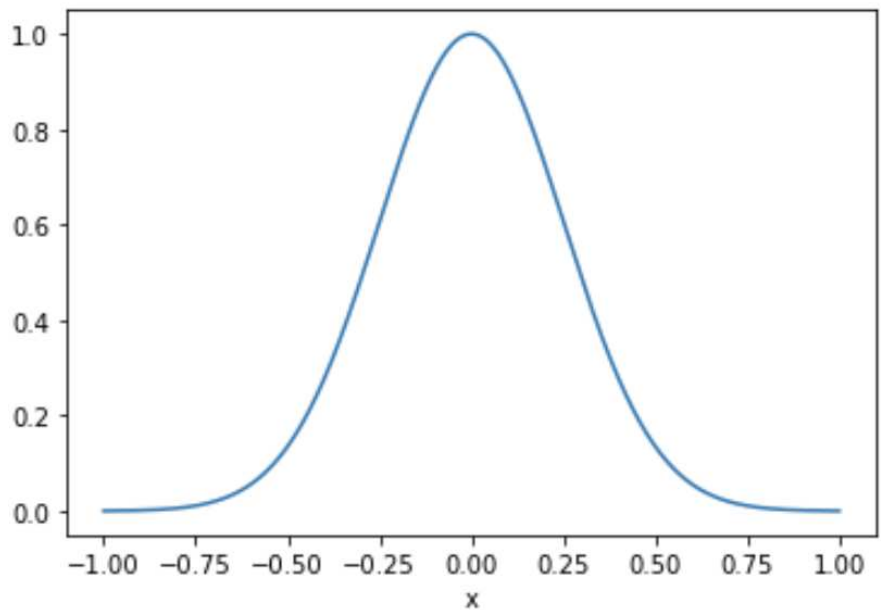}
\caption{Plot of \(\exp(-\frac{1}{2}(\frac{x}{l})^2) \) on \([-1,1]\), with \(l= \frac{1}{4}\).}
\end{figure}

\noindent Once again, pertaining to the above assumptions, the transition probabilities will also be estimated as
\begin{equation}
   \mathbb{P}(X_{n} = [x_{1},x_{2}] | X_{n-1} = [y_{1},y_{2}]) = \frac{\displaystyle \int_{x_{1}}^{{x_2}} \int_{{y_1}}^{{y_2}} \kappa(x,y)\;\mathrm{d}x \;\mathrm{d}y}{\displaystyle \int_{0}^{1} \int_{{y_1}}^{{y_2}}\kappa(x,y)\;\mathrm{d}x \;\mathrm{d}y} 
\end{equation}\\

\noindent Heuristically, this is used to determine how close two points have to be to influence each other significantly. This model allows for interiors of intervals to interfere with each other and make contributions to the total probability of a specific transition. Using figure \ref{fig: exp} as a guide, one notices that the length scale \(l\) is chosen in a fashion such that if the separation of two points is greater that half the length of the domain of \(\gamma\), then, their contribution to the probability becomes minimal.\\
\\It is clear that the farther apart two disjoint intervals are, one possible way to gauge this is using their Hausdorff distance, the probability given by the model will be expected to be less than if the intervals were close so that the mean separation between points in the intervals is within the 'support' of the kernel.\\
\\Also, by the mean value theorem for integrals, one obtains up to a constant of proportionality that for intervals \([x_1,x_2]\) and \([x_1', x_2']\) 

\[
\mathbb{P}(X_{n} = [x_{1},x_{2}] | X_{n-1} = [y_{1},y_{2}]) \sim (x_2-x_1) \displaystyle  \int_{{y_1}}^{{y_2}} \kappa(\xi_1,y) \;\mathrm{d}y\]

\[\mathbb{P}(X_{n} = [x_{1}',x_{2}'] | X_{n-1} =  [y_{1},y_{2}]) \sim (x_2'-x_1') \displaystyle  \int_{{y_1}}^{{y_2}} \kappa(\xi_2,y) \;\mathrm{d}y
\]

\noindent where \(\xi_1 = (x_1,x_2)\) and \(\xi_2 \in (x_1',x_2')\). If now one chooses the above intervals such that \(d(\xi_1, [y_1, y_2]) \approx d(\xi_2, [y_1, y_2])\), then the relative lengths of the intervals drives the relative behaviour of the probabilities of landing in said intervals. Thus, the modelling assumptions laid out above are adequately addressed by this model. \\

\noindent The transition matrix and transition diagram are as follows:
\begin{center}
\[\mathbf{P_{2}} = 
\begin{bmatrix}
& 0.84 & 0.07 & 0 & 0.09\\
& 0.50 & 0.15 & 0.002 & 0.348\\
& 0.43 & 0.16 & 0.002 & 0.408\\
& 0.32 & 0.158 & 0.002 & 0.52 \\
\end{bmatrix}
\]
\end{center}
Applying the exact same procedure as we did with the previous stochastic matrix, we obtain the following limiting distribution\\
\[
\mathbf{\pi_2} = \Big(0.7 \text{ \space  \space} 0.1 \text{ \space  \space} 0 \text{ \space  \space} 0.2 \Big)
\]
\begin{center}
\begin{tikzpicture}[->, >=stealth', auto, semithick, node distance=4cm]
\tikzstyle{every state}=[ fill=white,draw=black,thick,text=black,scale = 1.2]
\node[state]    (HM)                     {$HM$};
\node[state]    (SM)[above right of=HM]   {$SM$};
\node[state]    (LSM)[below right of=HM]   {$LSM$};
\node[state]    (EFSM)[below right of=SM]   {$EFSM$};
\path
(HM) edge[loop left]     node{$0.84$}         (HM)
    edge[bend left]     node{$0.07$}     (SM)
    edge[bend left,below]      node{$0.09$}      (EFSM)
    edge[bend left, below]    node{$0$}      (LSM)
(SM) edge[bend right]    node{$0.348$}           (EFSM)
    edge[bend left]     node{$0.50$}              (HM)
    edge[bend left]     node{$0.002$}              (LSM)
    edge[loop above]    node{$0.15$}              (SM)
(LSM) edge[bend right]    node{$0.408$}           (EFSM)
    edge[bend left]     node{$0.43$}              (HM)
    edge[loop below]    node{$0.002$}              (LSM)
    edge[bend left]     node{$0.16$}              (SM)
(EFSM) edge[loop right]    node{$0.52$}     (EFSM)
    edge[bend right,right]    node{$0.158$}(SM)
    edge[bend right]     node{$0.002$}      (LSM)
    edge[bend left,above]     node{$0.32$}   (HM);
\end{tikzpicture}
\end{center}

\section{Numerical Results}

\noindent In the simulation, the network will be sampled at constant intervals of length one unit of time, staying consistent with the first two Markov Models. Thus, in the above framework, we take \(\tau\) to be

\[
\mathbf{\tau} = \{0,1,2,\dots, T-1\}
\]
\noindent where \(T\) is some predefined constant. For the transition probabilities, a value of \(T = 1000\) will be used.

\begin{figure}[H]
\label{fig: gamma time series}
\includegraphics[width = \textwidth]{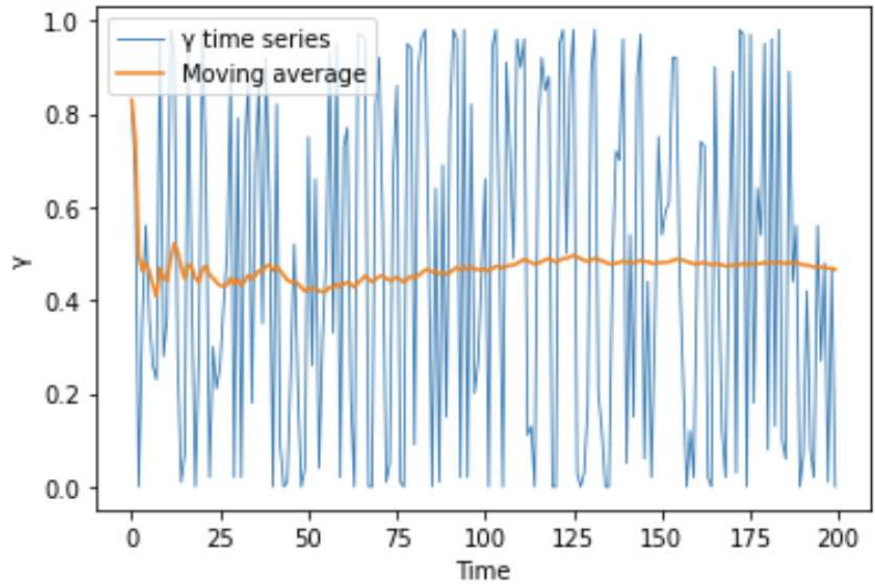}
\caption{Times series of \(\gamma \) from time with \(T = 200\) uniformly spaced samples, including the moving average}
\end{figure}

\noindent Below are the matrix of frequencies of each transition and the resulting transition probability matrix using \(T = 5000\) samples; though a time series for \(\gamma\) (see figure \ref{fig: gamma time series}) with \(T = 200\) is included for convenience. 

\begin{center}
\[\mathbf{N} = 
\begin{bmatrix}
&2977 & 205 & 0 & 690\\
&213 & 10 & 1 & 34\\
&8 & 1 & 0 & 6\\
&676 & 42 & 0 & 136 \\
\end{bmatrix}
\]
\end{center}

\begin{center}
\[\mathbf{P_{3}} = 
\begin{bmatrix}
& 0.77 & 0.05 & 0.0 & 0.18 \\
& 0.83 & 0.036 & 0.004 & 0.13 \\
& 0.53 & 0.07 & 0.0 & 0.4 \\
& 0.79 & 0.05 & 0.0 & 0.16 \\
\end{bmatrix}
\]
\end{center}
The chain represented by the above matrix is irreducible therefore we can apply the same techniques to derive a limiting distribution as we did for the previous two matrices.  We obtain:

\[\mathbf{\pi_3} = \Big(0.78 \text{ \space  \space} 0.05 \text{ \space  \space} 0 \text{ \space  \space} 0.17 \Big)
\]

\section{Likelihood Analysis}\label{log likelihood}

  \noindent Using the numerical model as a proxy for real data \(\mathbf{N}\), we will perform a likelihood analysis to compare the Markov models through their transition probability matrices \(\mathbf{P_{1}}, \mathbf{P_{2}}\). \\
 
 Formally, we consider the parameter space of Markov models 
 \[
 \Theta = \left\{\mathbf{P}\in \mathbb{R}^{4 \times 4}_{\geq 0} \bigg| \forall 0\leq i \leq3, \displaystyle \sum_{0\leq j \leq3}\mathbf{P}_{ij} = 1\right\}
 \]
 
 \noindent Now, the likelihoods of the models \(\mathbf{P_1}, \mathbf{P_2}\) given the time series for \(\mathbf{\gamma}\), yielding \(\mathbf{N}\) are computed using the Markov Property: 

\[
\mathcal{L}(\theta|N) = \displaystyle \prod_{0\leq i,j \leq 3} \mathbf{P}_{ij}^{\mathbf{N}_{ij}}\times\mathbb{P}(\gamma_{0} \in \Xi_{0} )
\]

\noindent we will see that the precise value of \(\mathbb{P}(\gamma_{0} \in \Xi_{0} )\) is irrelevant, so long as it is uniform across all models - an assumption we will make henceforth. Following \cite{davison2003statistical}, we define for a model \(\theta \in \Theta\), its Relative Ratio as follows:
\[
R\mathcal{L}(\theta|\mathbf{N}) = -\log \left(\frac{ \mathcal{L}(\mathbf{\theta}|\mathbf{N})}{\displaystyle \sup_{\theta \in \Theta}\mathcal{L}(\mathbf{\theta}|\mathbf{N})}\right)= -\log\left(\frac{ \mathcal{L}(\mathbf{\theta}|\mathbf{N})}{ \mathcal{L}(\mathbf{\hat{\theta}}|\mathbf{N})} \right)
\]
\[
 = -\left(\displaystyle \sum_{i,j} \mathbf{N}_{ij}\log(\mathbf{\theta}_{ij}) - \displaystyle \sum_{i,j} \mathbf{N}_{ij}\log(\mathbf{\hat{\theta}}_{ij})\right) \in [0,\infty]
\]

\noindent where \(\hat{\theta} = P_{3}\), a known result in likelihood optimisation. Note that if \(\mathcal{L}(\mathbf{\theta_1}|\mathbf{N})< \mathcal{L}(\mathbf{\theta_2}|\mathbf{N})\), then we have \(R\mathcal{L}(\theta_1|\mathbf{N}) > R\mathcal{L}(\theta_2|\mathbf{N}) \). We now compute the log-relative likelihoods of models \(\theta = \mathbf{P}_1, \mathbf{P}_2\):
 
 \begin{table}[h!]
 
\centering
\begin{tabular}{||c c||} 
 \hline
 \( R\mathcal{L}(\mathbf{P}_1|\mathbf{N}) \) & \( R\mathcal{L}(\mathbf{P}_2|\mathbf{N}) \) \\ [0.5ex] 
 \hline\hline
  157 & 512\\ [1ex] 
 \hline
\end{tabular}
\caption{Relative likelihoods of Markov models given data \(N\). \label{Tab:loglikelihood}}
\end{table}

\section{Conclusion}

\noindent Now that we have computed the limiting distributions from the three models discussed above, it is time to give them an interpretation. This will be achieved through the Ergodic theorem for Markov chains (see \cite{norris1998markov}), where for an irreducible and aperiodic homogeneous Markov chain \(\{X_n\}_{n\in\mathbb{N}}\) with limiting distribution \(\mathbf{\pi}\), with probability one the ratio counting the ratio of time spent in state \(i\)

\[
V_{i}(n) = \displaystyle \frac{1}{n}\sum_{k=1}^n\mathbf{1}_{\{X_{k}=i\}} \rightarrow \frac{1}{\pi_i}
\]
\noindent as \( n\rightarrow \infty\). \\

\noindent Applying the above result to our models which can be seen to satisfy the above conditions, the limiting distributions

\[
\mathbf{\pi_1} = \Big(0.73 \text{ \space  \space} 0.08 \text{ \space  \space} 0\text{ \space  \space} 0.19 \Big)\\
\]

\[
\mathbf{\pi_2} = \Big(0.7 \text{ \space  \space} 0.1 \text{ \space  \space} 0 \text{ \space  \space} 0.2 \Big)
\]

\[
\mathbf{\pi_3} = \Big(0.78 \text{ \space  \space} 0.05 \text{ \space  \space} 0 \text{ \space  \space} 0.17 \Big)
\]

\noindent are taken to measure as measuring the fraction of time spent in each optimal strategy for \(\gamma\) in the long run, where optimality is taken in the sense of \cite{https://doi.org/10.48550/arxiv.1805.08281}.\\

\noindent The log-likelihood analysis in section \ref{log likelihood}, the first model, namely  \(\mathbf{P_1}\) achieved a higher relative likelihood than the model \(\mathbf{P_2}\), due to the relative log likelihoods computed in table \ref{Tab:loglikelihood}. \\
\\We note that this observed difference with respect to the numerical data is due to the different theoretical premises they were derived from. For instance, the first model collapsed the intervals for \(\gamma\) into their midpoints, whereas the second model exploited the non linear interaction of all of the interiors of said intervals, provided by the kernel \(\kappa(x,y)\). Although, as discussed the paper, the qualitative features were broadly similar for they were meant to model the same underlying stochastic process \(\gamma\). \\

\noindent Moreover, we see that even if the attacker has a hashrate of 20\% in the Bitcoin network, the limiting distributions \(\mathbf{\pi_1}, \mathbf{\pi_2}, \mathbf{\pi_3}\) show that honest mining is strongly dominant in the long run, where it is used more than \(70\%\) of the time spent mining, as opposed to rogue mining strategies.

\section{Author Contribution Statement}

Y.P. conceived of the presented idea, namely the construction of Markov models for \(\gamma\). Y.P. developed the theoretical formalism for the first analytical model and performed the calculations of the transition matrices and limiting distributions.\\

\noindent P.T. conceived of the theoretical formalism of second analytical model and the numerical model. P.T. produced the code in the appendix to perform numerical simulations and performed the log-likelihood analysis of the analytical models using the numerical model as a benchmark.\\

\noindent Both authors discussed the results and contributed to the final manuscript.
\section{Appendix}
\subsection{Proof of Properties for Probability Models}
In this section we will only prove the first property for the first model. The remaining proof is similar and is left as an exercise to the reader. Let $X_{1} = [x_{1}, x_{2}]$, $X_{2} = [x_{3}, x_{4}]$ and $Y_{1} = [y_{1}, y_{2}]$. 
\[ \text{Let \space} \beta = \displaystyle\sum_{\xi \in \Xi}\displaystyle \int_{x_{1}}^{x_{2}} 1 - d_p([x_{1}, x_{2}],\xi) \;\mathrm{d}x \]
\[
d_{p}(X_{1}, Y) \geq d_{p}(X_{2}, Y)
\]
\[
1 - d_{p}(X_{1}, Y) \leq 1 - d_{p}(X_{2}, Y)
\]
\[
d(x_{1}, x_{2})(1 - d_{p}(X_{1}, Y_{1})) \leq d(x_{3}, x_{4}) (1 - d_{p}(X_{2}, Y_{1}))
\]
\[
(x_{2} - x_{1})(1 - d_{p}(X_{1}, Y_{1})) \leq (x_{4} - x_{3}) (1 - d_{p}(X_{2}, Y_{1}))
\]
\[
\int_{x_{1}}^{x_{2}} 1 - d_{p}(X_{1}, Y_{1})\;\mathrm{d}x \leq \int_{x_{3}}^{x_{4}} 1 - d_{p}(X_{2}, Y_{1}) \;\mathrm{d}x
\]
\[
\frac{1}{\beta} \cdot \int_{x_{1}}^{x_{2}} 1 - d_{p}(X_{1}, Y_{1})\;\mathrm{d}x \leq \frac{1}{\beta} \cdot \int_{x_{3}}^{x_{4}} 1 - d_{p}(X_{2}, Y_{1}) \;\mathrm{d}x
\]
\[
\therefore \mathbb{P}(X_{n} = [x_{1},x_{2}] | X_{n-1} = [y_{1},y_{2}]) \leq \mathbb{P}(X_{n} = [x_{3},x_{4}] | X_{n-1} = [y_{1},y_{2}])
\]

\newpage
\subsection{Numerical Model Implementation}

\begin{lstlisting}[language = Python, caption = Transition probabilities]
# Transitions constructed using hash rate of mining pool = 0.2
Transitions = {0:[0,0.675], 1: [0.675000001, 0.76], 2: [0.76000000001, 0.761], 3: [0.761000001, 1]}


P = [[0 for _ in range(4)] for _ in range(4)]

for n in range(T-1):
    if Transitions[0][0]<= y[1][n] <=Transitions[0][1] and Transitions[0][0]<=y[1][n+1] <=Transitions[0][1]:
        P[0][0]+=1
    elif Transitions[0][0]<=y[1][n] <=Transitions[0][1] and Transitions[1][0]<=y[1][n+1] <=Transitions[1][1]:
        P[0][1]+=1
    elif Transitions[0][0]<=y[1][n] <=Transitions[0][1] and Transitions[2][0]<=y[1][n+1] <=Transitions[2][1]:
        P[0][2]+=1
    elif Transitions[0][0]<=y[1][n] <=Transitions[0][1] and Transitions[3][0]<=y[1][n+1] <=Transitions[3][1]:
        P[0][3]+=1
        
    elif Transitions[1][0]<= y[1][n] <=Transitions[1][1] and Transitions[0][0]<=y[1][n+1] <=Transitions[0][1]:
        P[1][0]+=1
    elif Transitions[1][0]<=y[1][n] <=Transitions[1][1] and Transitions[1][0]<=y[1][n+1] <=Transitions[1][1]:
        P[1][1]+=1
    elif Transitions[1][0]<=y[1][n] <=Transitions[1][1] and Transitions[2][0]<=y[1][n+1] <=Transitions[2][1]:
        P[1][2]+=1
    elif Transitions[1][0]<=y[1][n] <=Transitions[1][1] and Transitions[3][0]<=y[1][n+1] <=Transitions[3][1]:
        P[1][3]+=1
        
    elif Transitions[2][0]<= y[1][n] <=Transitions[2][1] and Transitions[0][0]<=y[1][n+1] <=Transitions[0][1]:
        P[2][0]+=1
    elif Transitions[2][0]<=y[1][n] <=Transitions[2][1] and Transitions[1][0]<=y[1][n+1] <=Transitions[1][1]:
        P[2][1]+=1
    elif Transitions[2][0]<=y[1][n] <=Transitions[2][1] and Transitions[2][0]<=y[1][n+1] <=Transitions[2][1]:
        P[3][2]+=1
    elif Transitions[2][0]<=y[1][n] <=Transitions[2][1] and Transitions[3][0]<=y[1][n+1] <=Transitions[3][1]:
        P[2][3]+=1 
        
    elif Transitions[3][0]<= y[1][n] <=Transitions[2][1] and Transitions[0][0]<=y[1][n+1] <=Transitions[0][1]:
        P[3][0]+=1
    elif Transitions[3][0]<=y[1][n] <=Transitions[2][1] and Transitions[1][0]<=y[1][n+1] <=Transitions[1][1]:
        P[3][1]+=1
    elif Transitions[3][0]<=y[1][n] <=Transitions[2][1] and Transitions[2][0]<=y[1][n+1] <=Transitions[2][1]:
        P[3][2]+=1
    elif Transitions[3][0]<=y[1][n] <=Transitions[3][1] and Transitions[3][0]<=y[1][n+1] <=Transitions[3][1]:
        P[3][3]+=1




for i in range(4):
    a = sum(P[i])
    for j in range(4):
        P[i][j] = P[i][j]/a
        
#Transition matrix
P
\end{lstlisting}

\begin{lstlisting}[language = Python, caption = \(\tau \) strategy \(\gamma \) simulator including moving average]
from random import choices
T = 1000
T = [n for n in range(T)]

M = choices(list(range(1,100)),k = T)
N = choices(list(range(1,100)),k = T)
for i in N: 
    if N[i] == M[i]:
        v = list(range(1,100))
        v.remove(N[i])
        M[i] = choices(v,k = 1)[0]

y = stopping_time_simulator(T, N, M)
Y = [sum(y[1][:n])/n for n in range(1, T+1)]


plt.plot(T, y[1], linewidth = 0.8, label = "\gamma time series")
plt.xlabel("Time")
plt.ylabel("gamma")
plt.legend()
plt.plot(T, Y,  "Moving average")
plt.xlabel("Time")
plt.ylabel("gamma")
plt.legend()
plt.show()
\end{lstlisting}

\begin{lstlisting}[language=Python, caption=Graph Class]
import numpy as np
import matplotlib.pyplot as plt
from math import sqrt

class Graph():
 
    def __init__(self, vertices):
        self.V = vertices
        self.graph = [[0 for column in range(vertices)]
                      for row in range(vertices)]
 
    def printSolution(self, dist):
        print("Vertex \t Distance from Source")
        for node in range(self.V):
            print(node, "\t\t", dist[node])
            
    # A utility function to find the vertex with
    # minimum distance value, from the set of vertices
    # not yet included in shortest path tree
    def minDistance(self, dist, sptSet):
 
        # Initialize minimum distance for next node
        Min = 1e7
        min_index = 0
        # Search not nearest vertex not in the
        # shortest path tree
        for v in range(self.V):
            if dist[v] < Min and sptSet[v] == False:
                Min = dist[v]
                min_index = v
 
        return min_index
 
    # Function that implements Dijkstra's single source
    # shortest path algorithm for a graph represented
    # using adjacency matrix representation
    
    def dijkstra(self, src):
     
        dist = [1e7] * self.V
        dist[src] = 0
        sptSet = [False] * self.V
 
        for cout in range(self.V):
 
            # Pick the minimum distance vertex from
            # the set of vertices not yet processed.
            # u is always equal to src in first iteration
            u = self.minDistance(dist, sptSet)
 
            # Put the minimum distance vertex in the
            # shortest path tree
            sptSet[u] = True
 
            # Update dist value of the adjacent vertices
            # of the picked vertex only if the current
            # distance is greater than new distance and
            # the vertex in not in the shortest path tree
            for v in range(self.V):
                if (self.graph[u][v] > 0 and
                   sptSet[v] == False and
                   dist[v] > dist[u] + self.graph[u][v]):
                    dist[v] = dist[u] + self.graph[u][v]
 
        #self.printSolution(dist)
        return dist
    
    def Eigenvector_Centrality(self):
        # normalize starting vector
        x = dict([(n,1.0/self.V) for n in range(self.V)])
        s = 1.0/sum(x.values())
        for k in x:
            x[k] *= s
        Number_Nodes = self.V

        # make up to max_iter iterations
        max_iter = 50
        for i in range(max_iter):
            xlast = x
            x = dict.fromkeys(xlast, 0)

            # do the multiplication y = Cx
            # C is the matrix with entries
            Alpha = [xlast[k] for k in range(self.V)]
            C = [[0 for _ in range(self.V)] for _ in range(self.V)]
            for i in range(self.V):
                for j in range(self.V):
                    if self.graph[i][j] != 1E7:
                        C[i][j] = 1
            B = np.matrix(C).dot(Alpha)
            x = dict((n, B.item(n)) for n in range(self.V))

            # normalize vector
            try:
                s = 1.0/sqrt(sum(v**2 for v in x.values()))

            # this should never be zero?
            except ZeroDivisionError:
                s = 1.0
            for n in x:
                x[n] *= s

            # check convergence
            tol = 1E-5
            err = sum([abs(x[n]-xlast[n]) for n in x])
            if err < Number_Nodes*tol:
                return x
        return x
  
\end{lstlisting}

\begin{lstlisting}[language=Python, caption=Time Series Class]
class Node:
    import time
    def __init__(self, dataval=None, Time=time.time()):
        self.dataval = dataval
        self.time = Time
        self.nextval = None

class time_series():
    def __init__(self):
        self.headval = None

\end{lstlisting}

\begin{lstlisting}[language=Python, caption=Initialise Bitcoin Network]
def generate_blockchain(Number_Nodes):
    Blockchain = Graph(Number_Nodes)
    #adjacency matrix 
    W = [[0 for _ in range(Number_Nodes)] for _ in range(Number_Nodes)]
    
    for i in range(Number_Nodes-1):

        for j in range(i+1, Number_Nodes):
            for l in range(5):
                for m in range(l,6):
                    if (i in Intervals[l]) and (j in Intervals[m]):

                        if np.random.uniform(0,1)<0.1:

                            #if connection is not active - SimBlock Paper implementation
                            W[i][j] = 1E7
                            W[j][i] = 1E7
                        else:
                            #latency if connection is active -  SimBlock Paper implementaiton
                            mean = Region_Latency[l][m]
                            rand = np.random.poisson(mean)
                            shape = 0.2 * mean;
                            scale = mean - 5;
                            rand = int(scale / pow(np.random.uniform(0,1), 1.0 / shape))
                            W[i][j] = rand
                            W[j][i] = rand
            #count+=1
    Blockchain.graph = W
    return Blockchain


\end{lstlisting}

\begin{lstlisting}[language=Python, caption=Network \(\gamma\) time series simulator ]
def stopping_time_simulator(Tau, N, M):
    # give time series of gamma between two nodes on the blockchain sampled at times T
    gamma = []
    Number_Nodes = 100
    Network = time_series()
    network_init = generate_blockchain(Number_Nodes)
    Omega = network_init.Eigenvector_Centrality()        
    #bias parameter for activaiton in network:

    
    dist_N = network_init.dijkstra(N[0])
    dist_M = network_init.dijkstra(M[0])
    N_close = 0
    G = 0

    for i in set(range(Number_Nodes))-{N[0],M[0]}:
        #Effect of time interval on next iteration of network 
        if dist_N[i] < dist_M[i]:
            N_close += 1
    G = N_close/Number_Nodes
    gamma += [G]
    
    Network.headval = Node(network_init, Tau[0])
    pointer = Network.headval
    
    for n in range(1,len(Tau)):
        prevNetwork = pointer.dataval
        Alpha = gamma_dist(N[n],M[n],1, Tau[n]-Tau[n-1], prevNetwork)
        pointer.nextval = Node(Alpha[0], Tau[n])
        pointer = pointer.nextval
        gamma += Alpha[1]
    return (Network, gamma)
\end{lstlisting}

\begin{lstlisting}[language = Python, caption = Stochastic simulation model of Bitcoin Network]

\label{lst:Stochastic simulation Btc}

import numpy as np
from scipy.stats import skewnorm


#Latency distribution:

#Use heavy-tail skew distribution (skew-normal)

Regions_Distribution = {"NORTH_AMERICA": 0.3316, "EUROPE": 0.4998, "SOUTH_AMERICA":0.0090, "ASIA_PACIFIC":0.1177
                        , "JAPAN":0.0224,
                    "AUSTRALIA":0.0195}

Region_Latency = [[32, 124, 184, 198, 151, 189],
      [124, 11, 227, 237, 252, 294],
      [184, 227, 88, 325, 301, 322],
      [198, 237, 325, 85, 58, 198],
      [151, 252, 301, 58, 12, 126],
      [189, 294, 322, 198, 126, 16]]

Nodes_Region = [33,50,1,12,2,2]
Number_Nodes = 100
Nodes_Cumulative = np.cumsum(Nodes_Region)
Intervals = [list(range(Nodes_Cumulative[0]))]+[list(range(Nodes_Cumulative[i], Nodes_Cumulative[i+1])) for i in range(5)]


def gamma_dist(N,M,C, DeltaT, prevNetwork):
    Gamma = []
    for k in range(C):

        Blockchain = Graph(Number_Nodes)
        W = [[0 for _ in range(Number_Nodes)] for _ in range(Number_Nodes)]
        '''
        from math import comb
        K = comb(Number_Nodes,2)
        P = [np.random.uniform(-1,1) for _ in range(K)] 
        #M is the Correlation weight matrix - specific interpretation will be assigned later
        M = [[np.random.randint(1) for _ in range(K)] for _ in range(K)]
        #B = Bias , perhaps relate to bandwidth
        B = [np.random.uniform(0,1) for _ in range(K)]
        
        count = 0 
        '''
        #adjacency matrix 
        A = [[0 for _ in range(Number_Nodes)] for _ in range(Number_Nodes)]
        
        for i in range(Number_Nodes-1):
            for j in range(i+1, Number_Nodes):
                #previous network state influences connectivity
                if np.random.uniform(0,1) >= 0.1:
                    A[i][j] = 1
                    A[j][i] = 1
                                     
        
        Blockchain_unweighted = Graph(Number_Nodes)
        Blockchain_unweighted.graph = A

        #Eigenvector Centrality ranking of nodes - high score means node is connected to many highly connected nodes
        #The latency in the network is adjusted by the above centrality measure that accounts for topological properties of network
        
                
        Omega = Blockchain_unweighted.Eigenvector_Centrality()  
        
        for i in range(Number_Nodes-1):
            for j in range(i+1, Number_Nodes):
                for l in range(5):
                    for m in range(l,6):
                        if (i in Intervals[l]) and (j in Intervals[m]):
                            #previous network state influences connectivity
                            if prevNetwork.graph[i][j] != 1E7:
                                c = (Omega[i]+Omega[j])
                                if A[i][j] == 1:
                                    
                                    rand =  prevNetwork.graph[i][j] + prevNetwork.graph[i][j]*DeltaT*skewnorm.rvs(3*c, size = None)
                                    W[i][j] = rand 
                                    W[j][i] = rand
                                    
                                else: 
                                    W[i][j] = 1E7
                                    W[j][i] = 1E7  
                            else:
                                c = (Omega[i]+Omega[j])
                               
                                rand = Region_Latency[l][m] + Region_Latency[l][m]*DeltaT*skewnorm.rvs(3*c, size = None)
                                W[i][j] = rand
                                W[j][i] = rand
                                    


        Blockchain.graph = W
        dist_N = Blockchain.dijkstra(N)
        dist_M = Blockchain.dijkstra(M)
        N_close = 0
        gamma = 0
                
        for i in set(range(Number_Nodes))-{N,M}:
            
            if dist_N[i] < dist_M[i]:
                N_close += 1
        gamma = N_close/Number_Nodes
        Gamma.append(gamma)
    return (Blockchain, Gamma)
    


\end{lstlisting}

\printbibliography

\end{document}